\documentclass{emulateapj}

\newcommand{\Wlam}   {$W_{\lambda}$~}
\newcommand{\Wlams}  {$W_{\lambda}$s~}
\newcommand{\Wlamb}  {$W_{\lambda}$~}
\newcommand{\Wlamsb} {$W_{\lambda}$s~}
\newcommand{\Teff}   {$T_{\rm eff}$~}
\newcommand{\Teffb}  {$T_{\rm eff}$~}

\newcommand{\kap}    {$\kappa^1$~Cet~}

\shorttitle{$\kappa^1$~Ceti, an analog of the Sun when life arose on Earth}
\shortauthors{Ribas et al.}

\begin{document}

\title{Evolution of the solar activity over time and effects on
planetary atmospheres. II. $\kappa^1$~Ceti, an analog of the Sun when life
arose on Earth\footnote{Based on spectroscopic observations collected at the
Observat\'orio do Pico dos Dias (OPD), operated by the
Laborat\'orio Nacional de Astrof\'{\i}sica, CNPq, Brazil, at the
European Southern Observatory (ESO), within the ON/ESO and ON/IAG
agreements, under FAPESP project n$^{\circ}$ 1998/10138-8, and
with the Hubble Space Telescope.}}

\author{I. Ribas\altaffilmark{1}, G. F. Porto de Mello\altaffilmark{2}, 
L. D. Ferreira\altaffilmark{2}, E. H\'ebrard\altaffilmark{3,4}, 
F. Selsis\altaffilmark{3,4}, S. Catal\'an\altaffilmark{5}, 
A. Garc\'es\altaffilmark{1}, J. D. do Nascimento Jr.\altaffilmark{6},
and J. R. de Medeiros\altaffilmark{6}}

\altaffiltext{1}{Institut de Ci\`encies de l'Espai (CSIC-IEEC), Campus
UAB, Facultat de Ci\`encies, Torre C5, parell, 2a pl., E-08193
Bellaterra, Spain, \email{iribas,garces@ice.csic.es}} 
\altaffiltext{2}{Universidade Federal do Rio de Janeiro, Observat\'orio 
do Valongo, Ladeira do Pedro Ant\^onio 43, CEP: 20080-090, Rio de Janeiro, 
RJ, Brazil, \email{gustavo@astro.ufrj.br,leticia@astro.ufrj.br}} 
\altaffiltext{3}{Universit\'e de Bordeaux, Observatoire Aquitain des 
Sciences de l'Univers, 2 rue de l'Observatoire, BP 89, F-33271 Floirac 
Cedex, France, \email{franck.selsis,eric.hebrard@obs.u-bordeaux1.fr}}
\altaffiltext{4}{CNRS, UMR 5804, Laboratoire d'Astrophysique de Bordeaux, 
2 rue de l'Observatoire, BP 89, F-33271 Floirac Cedex, France}
\altaffiltext{5}{Centre for Astrophysics Research, Science and Technology 
Research Institute, University of Hertfordshire, Hatfield AL10 9AB, UK,
\email{s.catalan@herts.ac.uk}} 
\altaffiltext{6}{Departamento de F\'{\i}sica, Universidade Federal do Rio 
Grande do Norte, CEP: 59072-970, Natal, RN, Brazil, 
\email{dias,renan@dfte.ufrn.br}}

\begin{abstract}
The early evolution of Earth's atmosphere and the origin of life took 
place at a time when physical conditions at the Earth where radically 
different from its present state. The radiative input from the Sun was 
much enhanced in the high-energy spectral domain, and in order to model 
early planetary atmospheres in detail, a knowledge of the solar 
radiative input is needed. We present an investigation of the 
atmospheric parameters, state of evolution and high-energy fluxes of the 
nearby star $\kappa^1$~Cet, previously thought to have properties 
resembling those of the early Sun. Atmospheric parameters were derived 
from the excitation/ionization equilibrium of \ion{Fe}{1} and 
\ion{Fe}{2}, profile fitting of H$\alpha$ and the spectral energy 
distribution. The UV irradiance was derived from FUSE and HST data, and 
the absolute chromospheric flux from the H$\alpha$ line core. From 
careful spectral analysis and the comparison of different methods we 
propose for \kap the following atmospheric parameters: \Teff = 5665 
$\pm$ 30 K (H$\alpha$ profile and energy distribution), $\log g = 4.49 
\pm 0.05$ dex (evolutionary and spectroscopic) and $[Fe/H]=+0.10 
\pm0.05$ (\ion{Fe}{2} lines). The UV radiative properties of \kap 
indicate that its flux is some 35\% lower than the current Sun's between 
210 and 300 nm, it matches the Sun's at 170 nm and increases to at least 
2--7 times higher than the Sun's between 110 and 140 nm. The use of 
several indicators ascribes an age to \kap in the interval $\sim$ 
0.4--0.8~Gyr and the analysis of the theoretical HR diagram suggests a 
mass $\sim$ 1.04~M$_\odot$. This star is thus a very close analog of the 
Sun when life arose on Earth and Mars is thought to have lost its 
surface bodies of liquid water. Photochemical models indicate that the 
enhanced UV emission leads to a significant increase in 
photodissociation rates compared with those commonly assumed of the 
early Earth. Our results show that reliable calculations of the chemical 
composition of early planetary atmospheres need to account for the 
stronger solar photodissociating UV irradiation. 
\end{abstract}


\section{Introduction}

The irradiation from the parent star is, by far, the most important 
source of energy in planetary atmospheres. Most of the physical and 
chemical properties of the atmosphere of a planet are largely driven by 
the stellar input, which ultimately determines its structure, 
composition and, even, its mere existence. Most of the radiation emitted 
by the Sun comes from the photosphere and, at the solar effective 
temperature today (${T_{\rm eff}}_{\odot} = 5780$ K), it is the dominant 
source at wavelengths above 170 nm. The solar photospheric flux is quite 
stable over short timescales (decades) and it only suffers variations 
driven by sunspots and faculae with an amplitude below 0.2--0.3\% peak 
to peak \citep{froehlichlean2004}. Over long timescales, the variations 
have been larger and are related to the nuclear evolution of the Sun. 
Model predictions indicate that the young ZAMS Sun could have had a 
luminosity some 35\% lower than today \citep[unless a scenario of heavy 
mass loss is considered;][]{sackmannboothroyd2003}.

The energetic end of the solar spectrum (i.e., below 170 nm) is 
dominated by the emissions from high-temperature plasma in the 
chromosphere, transition region and corona. Such high-energy fluxes are 
strongly variable over short- and mid-term timescales because of flare 
events, rotational modulation, activity cycles, etc. 
\citep[e.g.,][]{leanetal1997}. Over longer timescales, several past 
studies \citep{ZW82,A97} established that the young Sun's high-energy 
emissions were possibly up to several orders of magnitude stronger than 
currently.

The Sun in Time programme \citep[hereafter Paper 
I]{dorrenguinan1994,ribasetal2005} focused on a small sample of 
carefully-selected and well-studied stellar proxies that represent key 
stages in the evolution of the Sun. This approach allowed the study in 
the X-ray, EUV, and FUV domains, where the variations are of one order 
of magnitude or more. However, the stellar proxy technique becomes 
increasingly uncertain in the UV since the stars are not perfect matches 
to the Sun (their masses are within 10\% of 1~M$_\odot$) and the 
expected flux variations are in the range of tens of a percent. For this 
reason, most of the Sun in Time stars employed in Paper I cannot be 
reliably used to infer the solar flux evolution at wavelengths longer 
than about 130 nm. But at the same time, the wavelength interval between 
130 and 200 nm is an essential energy input to planetary atmospheres 
since it drives most of the photochemical reactions 
\citep[e.g.,][]{CLA82,CLA83}. In addition, its impact on living 
organisms must also be considered, although it is unlikely that photons 
in this range could penetrate a dense atmosphere \citep{cockell2000}.

Among the solar proxies studied in the Sun in Time, \kap (HD 20630, HIP 
15457, $d=9.16$~pc, $V=4.84$) stands out as potentially having a mass 
very close to solar and a young age (see Paper I). This could be a very 
good analog of the Sun at the critical time when life is thought to have 
originated on Earth 3.8 Gyr ago, at the start of the Archean epoch 
\citep{mojzsisetal1996}. It is also about the time when Mars lost its 
liquid water inventory at the end of the Noachian epoch some 3.7 Gyr ago 
\citep{jakoskyphillips2001}. For these reasons, \kap deserves attention 
as a possible precise match to the young Sun, thus providing information 
on the radiation environment that determined the properties and chemical 
composition of the planetary atmospheres. \citet{cnossenetal2007} 
presented a study based on \kap with a goal set on assessing the 
biological implications of the high-energy radiations. The authors did 
not use a real high-energy spectrum but generated synthetic data using 
plasma models from an inferred emission measure distribution.

In this paper we carry out an in-depth analysis of $\kappa^1$~Cet, 
including its radiative properties, chemical abundances, atmospheric 
parameters and state of evolution, with the ultimate goal of 
understanding the Sun's UV emissions at a critical time in the past. The 
paper is organized as follows. In \S 2 we describe the spectroscopic 
analysis and derivation of atmospheric parameters, along with its Li 
abundance. The high-energy radiative processes and magnetic activity of 
\kap are discussed in the context of active stars in \S 3. In \S 4 we 
determine its mass and state of evolution, and in \S 5 we employ its 
properties to discuss the young Sun in the frame of photodissociation 
calculations. Our conclusions are drawn in \S 6.

\section{Spectroscopic atmospheric parameters}

\subsection{Visible spectra}

Visible spectroscopic observations were performed with two setups. The 
Coud{\'e} spectrograph of the 1.60-m telescope of Observat{\'o}rio do 
Pico dos Dias (OPD, Bras{\'o}polis, Brazil), operated by Laborat{\'o}rio 
Nacional de Astrof\'{\i}sica (LNA/CNPq), was used to obtain spectra of 
\kap and the Sun (represented by Moon spectra) in a 150 \AA\, spectral 
range, centered in the H$\alpha$ line, with R = 20\,000 and S/N ratios 
in excess of 300. \kap was observed in 2008, and the Moon spectra were 
acquired in a series of runs between 1994 and 2004, to ensure the 
absence of long-term systematics.

Additionally, the FEROS spectrograph \citep{kauferetal99} was used in 
2000 to obtain spectra of \kap and Ganymede (as a solar template), at a 
nominal resolution of R = 45\,000. The S/N ratio per resolution element 
was 800 and 1000, respectively, for \kap and the Sun, between 
$\lambda$$\lambda$5600--6960, being about half as much for the 
$\lambda$$\lambda$4500--5500 range. As \kap is very solar-like in its 
properties, the Sun is the natural choice as the standard star of a 
differential analysis. In this approach, systematic errors in line 
measurement, atmospheric modelling and the possible presence of 
Non-Local Thermodynamic Equilibrium (NLTE) effects tend to be minimized.

Data reduction was carried out by standard procedures using 
IRAF{\footnote {{\it Image Reduction and Analysis Facility} (IRAF) is 
distributed by the National Optical Astronomical Observatories (NOAO), 
which is operated by the Association of Universities for Research in 
Astronomy (AURA), Inc., under contract to the National Science 
Foundation (NSF).}}. Extreme care was taken in the continuum 
normalization to guarantee equivalent width (hereafter \Wlams) 
measurements as free from systematic effects as possible. This meant 
chosing specific wavelength intervals by comparing, window by window, 
\kap with the solar template (Ganymede) and the Solar Flux Atlas 
\citep{kuruczatlas} using the Utrecht spectral line compilation 
\citep{mooreetal1966}. For each window, the line strengths were checked 
to see if the wings of neighboring lines might be causing different 
depressions in the spectra. All windows considered suspect were not 
used. The remaining windows were then adjusted with low order 
polynomials. The spectral windows were always simultaneously normalized 
(Sun and $\kappa^1$~Cet) to minimize the possibility of errors. The 
\Wlams of \ion{Fe}{1} and \ion{Fe}{2} lines were measured by Gaussian 
fits.

A series of sanity checks were performed on the measured $W_{\lambda}$s. 
Saturated lines were eliminated by a 2$\sigma$ clipping on the relation 
of reduced width $W_{\lambda}/\lambda$ with line depth, and no lines 
were measured beyond the linearity limit. Also, no trend is expected in 
the relation of the line full-width-half-maximum and reduced width, 
since the line widths are essentially set by the instrumental profile. 
For stronger lines the inability of gaussian fits to correctly reproduce 
the line wings leads to an artificial increase of the line FWHM with 
reduced width, and no lines were measured beyond this limit. The final 
line selection comprised 90 \ion{Fe}{1} and 12 \ion{Fe}{2} lines. We 
follow the same procedure discussed in detail by 
\citet{portodemelloetal2008}. The measured \Wlams were corrected by 
+3.6\% to bring them onto a system compatible with the Voigt-fitted 
solar \Wlams of \citet{meylanetal1993}. The reason for the correction is 
both a low level of scattered light in the spectrograph ($\lesssim$~2\%) 
and also the inability of Gaussian fits to fully represent observed line 
profiles.

\subsection{Spectroscopic analysis}

We derived solar gf-values for the \ion{Fe}{1} and \ion{Fe}{2} spectral 
lines from a LTE, 1-D, homogeneous and plane-parallel solar model 
atmosphere from the NMARCS grid 
\citep{gustafssonetal2008,edvardssonetal1993}. The adopted parameters 
for the Sun were \Teff = 5780 K, $\log g = 4.44$, $[Fe/H]=+0.00$ and 
$\xi_t = 1.00$~km~s$^{-1}$, and we employed the \Wlamsb measured off the 
Ganymede spectra, corrected to the Voigt scale of 
\citet{meylanetal1993}. The adopted solar absolute abundances are those 
of \citet{asplundetal2009}.

The spectroscopic atmospheric parameters of \kap were determined by the 
simultaneous satisfaction of the excitation and ionization equilibria of 
\ion{Fe}{1} and \ion{Fe}{2}. \Teff was obtained by forcing the 
\ion{Fe}{1} line abundances to be independent of their excitation 
potential. Surface gravity was determined by forcing the lines of 
\ion{Fe}{1} and \ion{Fe}{2} to agree to the same abundance. The 
microturbulence velocity $\xi_t$ was set by forcing the lines of 
\ion{Fe}{1} to be independent of their $W_{\lambda}$s. The Fe abundance 
$[Fe/H]$ (we use throughout the notation $[A/B] = \log N(A)/N(B)_{\rm 
star} - \log N(A)/N(B)_{\rm Sun}$, where N denotes the number abundance) 
is automatically obtained at the end of the iteration, and this solution 
is unique for a given set of gf values, \Wlams and model atmospheres. 
Here we also used the same procedure as \citet{portodemelloetal2008}. 
The \ion{Fe}{1} lines span from 0 to 5 eV in excitation potential, and 
from 10 to 115 m\AA\, in $W_{\lambda}$, allowing for an internally very 
precise solution of the excitation and ionization equilibrium.

\begin{figure}
\epsscale{1.0}
\plotone{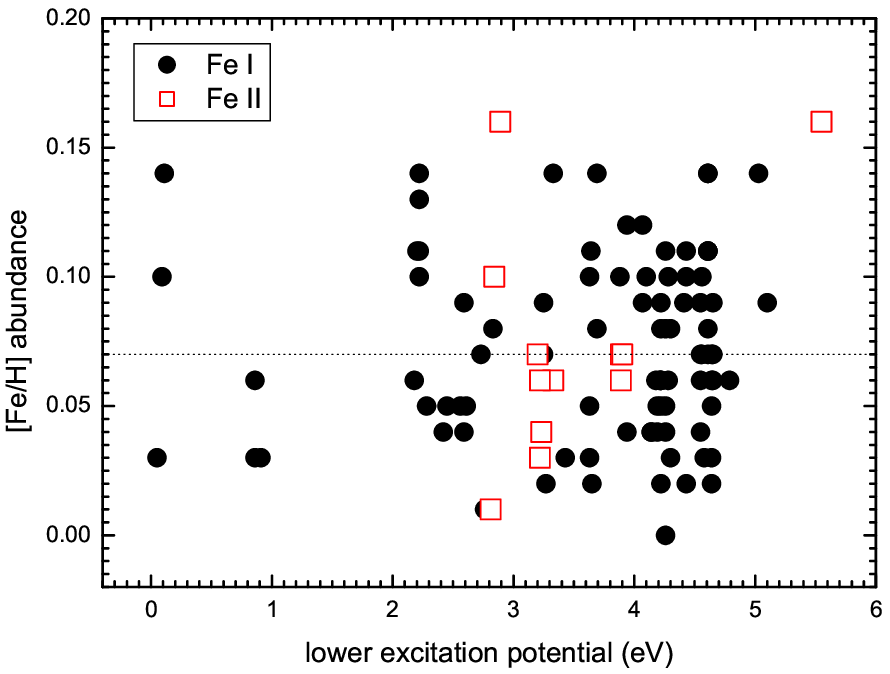}
\caption{The excitation \& ionization equilibrium of \ion{Fe}{1}
and \ion{Fe}{2} lines for \kap.} \label{excitation}
\end{figure}

The formal standard error in the spectroscopic \Teff was determined from 
the 1$\sigma$ uncertainty of the slope of the linear regression in the 
$[Fe/H]$ vs. $\chi$ diagram, yielding the \Teffb variation admissible at 
the 1$\sigma$ level. For the microturbulence velocity, the same 
procedure provides the 1$\sigma$ microturbulence uncertainty in the 
$[Fe/H]$ vs. \Wlamb diagram. For the metallicity $[Fe/H]$, we adopt the 
standard deviation of the distribution of abundances derived from the 
\ion{Fe}{1} lines, which is larger than the errors in $[Fe/H]$ due to 
\Teff, $\xi$ and \Wlam errors. The error of the spectroscopic $\log g$ 
is estimated by evaluating the variation in this parameter which 
produces a disagreement of 1$\sigma$ between the abundances of 
\ion{Fe}{1} and \ion{Fe}{2}. The spectroscopic results thus determined 
for \kap are: \Teff = 5780 $\pm$ 30 K, $\log g = 4.48 \pm 0.10$ dex, 
$[Fe/H] = +0.07 \pm 0.04$ dex (for the \ion{Fe}{2} lines, $\sigma = 
0.05$ dex). The excitation \& ionization equilibrium solution as a 
function of excitation potential is shown in Fig.~\ref{excitation} for 
all \ion{Fe}{1} and \ion{Fe}{2} lines.

\begin{deluxetable}{lcccc}
\tabletypesize{\footnotesize}
\tablecaption{Atmospheric parameters of \kap \label{tab:param}}
\tablewidth{0pt}
\tablehead{
\colhead{Authors} & \colhead{\Teff} & \colhead{$\log g$} & 
\colhead{$[Fe/H]$} & \colhead{Method\tablenotemark{a}}
}
\startdata
Cayrel de Strobel       &      &         &                 \\
\& Bentolila (1989) & 5630 & 4.50 & $+$0.04 & phot., H$\alpha$\\
\citet{pasquinietal1994} & 5675 & 4.35 & $-$0.01 & phot. \\
\citet{ottmanetal1998} & 5680 & --- & --- & H$\alpha$, H$\beta$ \\
\citet{gaidosgonzalez2002} & 5747 & 4.53 & $+$0.11 & exc. \& ion. \\
\citet{barklemetal2002} & 5710 & --- & --- & H$\alpha$,H$\beta$ \\
\citet{heiterluck2003} & 5750 & 4.55 & $+$0.05 & line depths \\
\citet{allendeprietoetal2004} & 5564 & 4.52 & $-$0.06 & phot. \\
\citet{luckheiter2006} & 5700 & 4.55 & $+$0.05 & exc. \& ion. \\
This work & 5780 & 4.48 & $+$0.07 & exc. \& ion. \\
This work & 5685 & --- & --- & phot. \\
This work & 5645 & --- & --- & H$\alpha$ \\
& & & & \\
{\bf ADOPTED} & {\bf 5665} & {\bf 4.49} & {\bf $+$0.10} & see text \\
\enddata
\tablenotetext{a}{Method: ``phot.'' stands for photometric derivations of 
$T_{\rm eff}$, ``H$\alpha$'' or ``H$\beta$''; for the fitting of 
theoretical profiles to the Balmer lines, ``line depths'' for the 
calibration of \Teff and line depth ratios of metal lines, and ``exc. \& 
ion.'' for the satisfaction of the excitation and ionization equilibrium 
with \ion{Fe}{1} and \ion{Fe}{2} lines.}
\end{deluxetable}

\begin{figure*}
\epsscale{1.0}
\plottwo{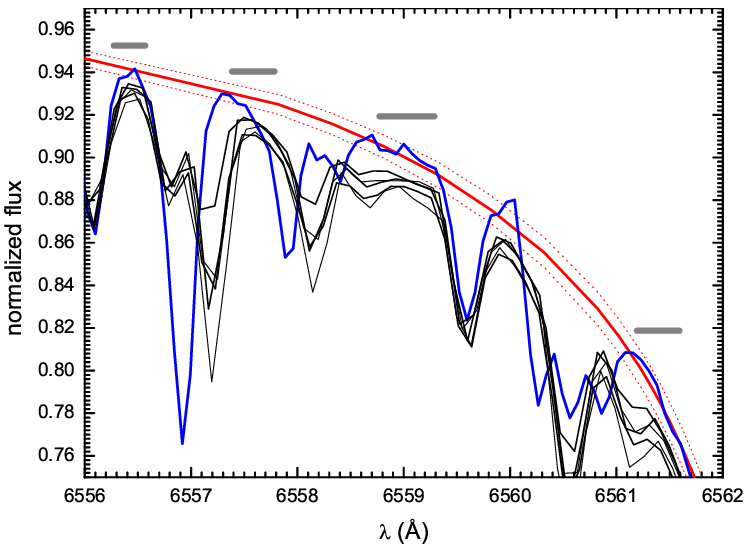}{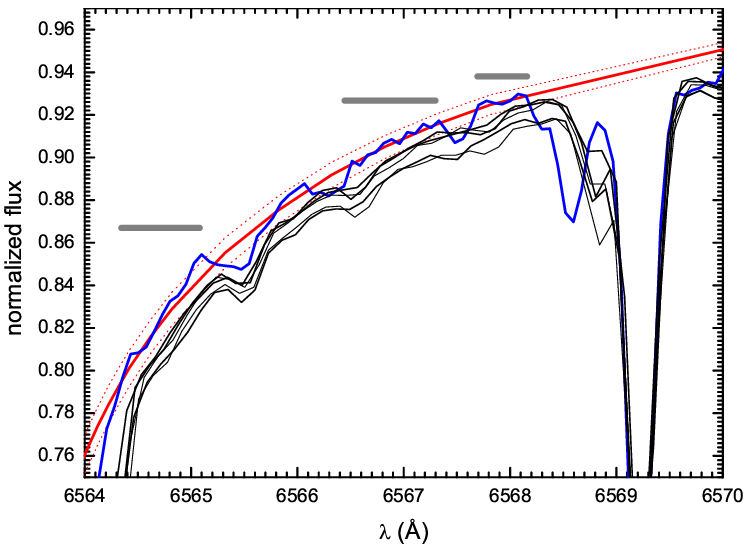}
\caption{{\em Left:} The blue wing of the H$\alpha$ profile of 
$\kappa^1$~Cet, plotted with five representative spectra of the Moon, 
observed from 1995 to 2002, and three theoretical models, centered at 
\Teff = 5645 K and spaced by 50 K. The gray horizontal bars denote the 
positions of the stellar profile regions free from telluric and metal 
lines. {\em Right:} Same as above for the red wing.} \label{halpha-wings}
\end{figure*}

An additional \Teff was determined by fitting the observed wings of 
H$\alpha$, following closely the method of \citet{lyraportodemello2005}. 
This procedure is shown in Fig. \ref{halpha-wings}. Here we varied from 
\citet{lyraportodemello2005} routine by selecting manually individual 
flux points in the blue and red wings of the observed \kap H$\alpha$ 
spectrum, free of telluric lines and photospheric line perturbations, in 
a total 21 and 43 points, respectively, for the blue and red wings. The 
\Teff values derived from each wing agree within 5 K, and a mean value 
\Teff = 5645 $\pm$ 40 K. This error refers exclusively to the dispersion 
of \Teff values attributed to the fitted profile data points.  A more 
formal estimate of the uncertainty of this procedure 
\citep{lyraportodemello2005}, taking into account errors in the input 
atmospheric parameters and the continuum normalization, leads to an 
error of 50 K.

\subsection{Effective temperature from the spectral energy
distribution}

The photometric \Teff of \kap was calculated using a variety of methods. 
Firstly, we employed intermediate-band Str\"omgren photometry collected 
from the GCPD database \citep{mermilliodetal1997}, which lists mean 
indices of $V = 4.850 \pm 0.008$, $(b-y) = 0.419 \pm 0.003$, $m_1 = 
0.235 \pm 0.005$, $c_1 = 0.307 \pm 0.003$, and $\beta = 2.595 \pm 
0.016$. We used the photometric grids of R. Napiwotzki (priv. comm.), 
which are based on ATLAS9 model atmospheres. The determination of the 
\Teff was done by considering the 1$\sigma$ uncertainties in the color 
indices and following a Monte Carlo procedure. From 1000 realizations we 
obtained a mean value and its corresponding standard deviation of \Teff 
= 5650 $\pm$ 125~K. The error bar only reflects the random uncertainty 
and does not account for systematic contributions. The Str\"omgren 
photometry also suggests roughly solar chemical composition.

A more accurate \Teff determination was obtained from the IRSED method 
of \citet{masanaetal2006}. This is based on the use of (2MASS) near-IR 
photometry and the fit of the spectral energy distribution with stellar 
atmosphere models. The calibration of \citet{masanaetal2006} takes into 
account possible systematic trends and applies corrections based on a 
sample of solar analogs. The photometry of \kap present in the 2MASS 
catalog is severely saturated but, fortunately, this star was selected 
as an ESO standard and numerous measurements are listed in 
\citet{bouchetetal1991}. From the ESO IR photometry ($J_{\rm ESO} = 
3.673 \pm 0.017$, $H_{\rm ESO} = 3.362 \pm 0.013$, $K_{\rm ESO} = 3.282 
\pm 0.013$), we applied transformations in \citet{carpenter2001} to 
obtain photometry in the 2MASS system ($J_{\rm 2MASS} = 3.605 \pm 
0.028$, $H_{\rm 2MASS} = 3.348 \pm 0.031$, $K_{\rm 2MASS} = 3.239 \pm 
0.018$). Using this photometry the IRSED method yields an effective 
temperature of \Teff = 5685 $\pm$ 45~K when assuming solar values for 
$\log g$ and $[Fe/H]$. The dependence of the result on these adoptions 
is very weak, and \kap does have these parameters very similar to the 
Sun's, as already shown.

\subsection{Systematic offset between photometric, H$\alpha$ and
spectroscopic \Teff values}

The three \Teff values derived in the present work are not in agreement 
when the standard errors are considered. The spectroscopic \Teff is 
$\sim$110 K higher than the H$\alpha$ and photometric ones, which are in 
close agreement. When one considers the results of recent spectroscopic 
analyses of $\kappa^1$~Cet, all of them based on high-quality, 
high-resolution spectra (Table~\ref{tab:param}), an interesting pattern 
emerges. Analyses based on photometric and Balmer line methods to derive 
\Teff cluster very systematically in low values around \Teff = 5650 K, 
while those relying on the excitation and ionization equilibria of Fe, 
or line depth ratios, cluster around \Teff = 5750 K. A recent discussion 
on a possible offset between the spectroscopic \Teff scale on one side, 
and Balmer line and photometric ones on the other, the former being the 
hotter, has been given by \citet{portodemelloetal2008}. Authors do not 
generally agree whether this offset is due to NLTE effects in cool 
stars, in the sense that for stars with \Teff $\sim$ 5000 K the offsets 
are large, while for \Teff $\sim$ 6000 K good agreement is found 
\citep{ramirezetal2007}: this discrepancy is revealed in this case 
either by a disagreement between chemical abundances derived from 
different lines of the same species (atomic or molecular) or by the 
non-realization of the \ion{Fe}{1}/\ion{Fe}{2} ionization equilibrium. 
Another interpretation is the presence of high chromospheric activity, 
an enhanced non-local UV radiation field and a resulting photospheric 
overionization \citep{schuleretal2006}. In a classical spectroscopic 
analysis, forcing agreement between \ion{Fe}{1} and \ion{Fe}{2} 
abundances under LTE, but in the presence of an overionizing radiation 
field, leads naturally to a higher $T_{\rm eff}$. As will be seen below, 
\kap is a very active star, with a chromospheric flux at the H$\alpha$ 
core only slightly lower than a typical Hyades solar-type star, as well 
as a heightened UV flux with respect to the Sun below $\lambda$1600 
\AA,\, besides a very high X-ray luminosity. Since \ion{Fe}{2} is 
essentially insensitive to NLTE effects for \Teff values similar to that 
of the Sun \citep{theveninidiart1999}, the most reliable determination 
of the Fe abundance in this instance is that due to the \ion{Fe}{2} 
lines, adopting the photometric and H$\alpha$ profile $T_{\rm eff}$.

Accepting the likely presence of NLTE effects in $\kappa^1$~Cet, we 
regard the following values as the most likely atmospheric parameters: 
\Teff = 5665 $\pm$ 50 K, as a straight average of the \Teff values 
derived from photometry and the H$\alpha$, $\log g = 4.49 \pm 0.10$ and 
$[Fe/H] = +0.10 \pm 0.05$ dex, the latter due exclusively from the 
\ion{Fe}{2} lines. Note that our metallicity value is in very good 
ageement with recent determinations \citep[e.g.,][]{valentifischer2005}.

\section{Magnetic activity and age}

\kap has always been recognized as a magnetically active star, with 
levels typical of a young solar analog. A space-based photometric study 
using the MOST satellite \citep{walkeretal2007} provided a precise value 
for its equatorial rotation period of 8.77 days, which is some 3 times 
faster than that of today's Sun. This gives rise to a significantly 
enhanced magnetic dynamo and consequently stronger 
magnetically-generated phenomena. The magnetic properties of 
$\kappa^1$~Cet, together with spectroscopic diagnostics, can be used to 
place constraints on the age of the star.

\subsection{H$\alpha$ absolute chromospheric flux}

\begin{figure}
\epsscale{1.0}
\plotone{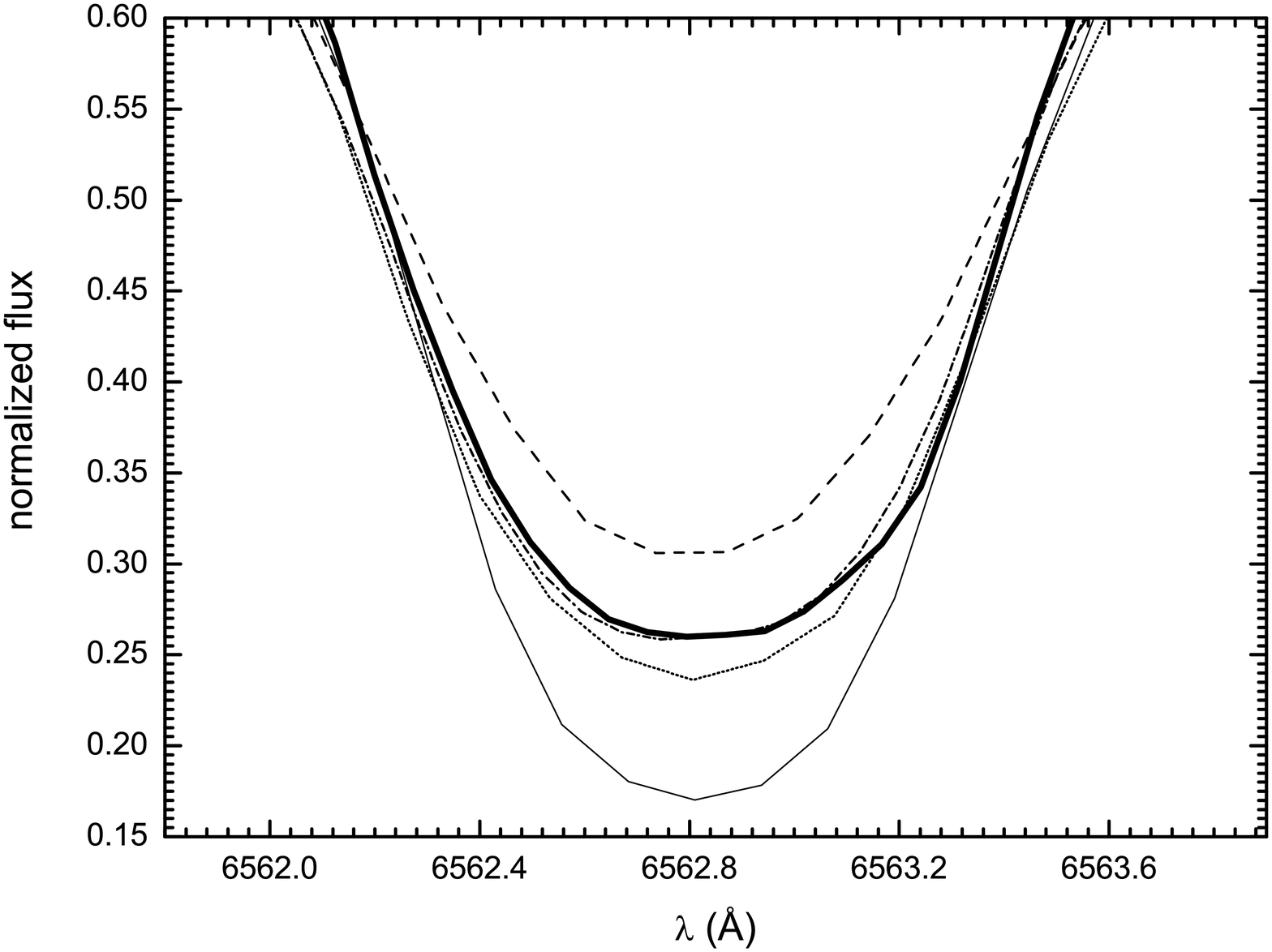}
\caption{The H$\alpha$ line core of $\kappa^1$~Cet (thick full line), the 
Sun (thin full line), and mean spectra of representative stars of the Pleiades
(dashed line) and Hyades (dotted line) clusters, and the Ursa Major moving
group (dash-dotted line).} 
\label{halpha-core}
\end{figure}

Absolute chromospheric radiative losses at the core of H$\alpha$ were 
derived by \citet{lyraportodemello2005} for a large sample of solar-type 
stars, including stars from Pleiades and Hyades clusters, and the Ursa 
Major kinematic group. The average fluxes (and dispersions) of three 
Pleiades stars, seven UMa Group stars and seven Hyades stars are, 
respectively, 15.0 $\pm$ 1.9, 9.5 $\pm$ 2.9 and 7.2 $\pm$ 1.2 (units are 
$10^5$~erg~cm$^{-2}$~s$^{-1}$~\AA$^{-1}$), in a clear age progression. 
\citet{lyraportodemello2005} estimate their H$\alpha$ flux uncertainty 
(probably underestimated because it does not account for errors in the 
photometric calibration of the fluxes) at 0.5 in the same units. In 
comparison to more classical spectroscopic indicators of chromospheric 
losses such as the H \& K lines of \ion{Ca}{2} \citep{pasquini1992}, 
H$\alpha$ fluxes are more subject to observational uncertainty, but are 
also less sensitive to activity cycle phase and rotational modulations, 
as well as transient phenomena \citep{lyraportodemello2005}. They should 
therefore be representative of the average level of stellar activity.

In Fig.~\ref{halpha-core}, mean spectra of the stellar groups, \kap and 
the Sun are overplotted. It is apparent that the mean chromospheric 
filling at the H$\alpha$ core is much higher for the Pleiades, and 
similar for the UMa group and the Hyades. The appearance of the core 
flux spectra does not translate directly to the chromospheric fluxes, 
since the core flux is integrated in a 1.7~\AA-wide window and \Teff 
differences among the group members are not negligible. Considering the 
dispersions of the mean H$\alpha$ core fluxes of the Hyades cluster and 
the UMa group, nearly the same mean H$\alpha$ activity level can be 
ascribed to them, and \kap is therefore compatible with their age range. 
Ages for these stellar groups are 0.1 Gyr for the Pleiades 
\citep{schilbachetal1995}, 0.63 Gyr for the Hyades 
\citep{perrymanetal1998} and $\sim$0.5 Gyr for the UMa Group 
\citep{kingetal2003}. Applying the flux calibration of 
\citet{lyraportodemello2005} to our H$\alpha$ spectrum leads to 
$7.3\cdot 10^5$~erg~cm$^{-2}$~s$^{-1}$~\AA$^{-1}$, which places \kap at 
the same flux level than an average Hyad, and below the flux level of an 
average member of the Ursa Major Group. The age calibration of 
\citet{lyraportodemello2005} yields 0.65 Gyr for $\kappa^1$~Cet, taking 
its chromospheric H$\alpha$ flux at face value.

\subsection{Lithium abundance}

\begin{figure}
\epsscale{1.0}
\plotone{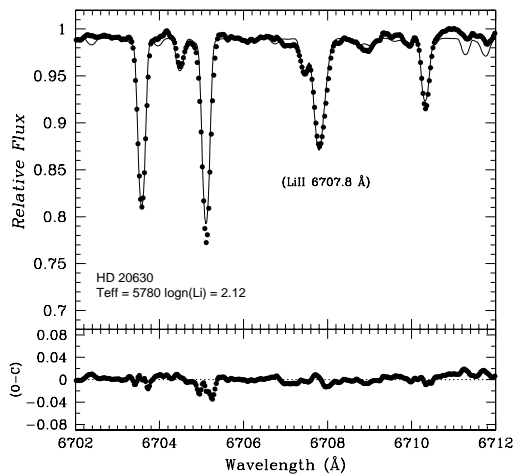}
\caption{Spectral synthesis of the $\lambda$6707 \ion{Li}{1} line of
$\kappa^1$~Cet: the purely spectroscopic set of atmospheric parameters was
used.} \label{lithium-5780}
\end{figure}

The lithium abundance of \kap was derived from the \ion{Li}{1} resonance 
transition at $\lambda$6707. A synthetic spectrum was fitted to the 
FEROS spectrum, for two sets of atmospheric parameters: the purely 
spectroscopic solution, \Teff = 5780~K, $\log g = 4.48$~dex, $[Fe/H] = 
+0.07$ and $\xi$ = 1.21 km~s$^{-1}$, and the photometric/H$\alpha$ 
solution, \Teff = 5665~K, $\log g = 4.49$ dex, $[Fe/H] = +0.10$ and 
$\xi$ = 1.20 km~s$^{-1}$. Model atmospheres were interpolated in the 
Kurucz grid \citep{kurucz_CD_1993} and the synthetic spectra were 
calculated with the MOOG routine \citep{sneden_1973_MOOG}. The FEROS 
instrumental broadening profile was set at 0.07 \AA. The synthesis of 
\ion{Fe}{1} lines in the $\lambda$6707 vicinity provides a projected 
rotational velocity $v \sin i = 5 \pm 1$~km~s$^{-1}$. For the 
spectroscopic atmospheric parameters, we derived $\log N({\rm Li}) = 
2.12$ (Fig.~\ref{lithium-5780}), in the usual scale where $\log N({\rm 
H}) = 12.00$. This value is in very good agreement with the 
determinations of \citet{luckheiter2006} and \citet{pasquinietal1994}, 
respectively, $\log N({\rm Li}) = 2.04$ and 2.13.

This Li abundance places $\kappa^1$~Cet, for the corresponding $T_{\rm 
eff}$, in very good agreement with the Li sequence for the Praesepe and 
Hyades clusters as determined by \citet{soderblometal1993}. The two 
clusters are thought to be coeval at $\sim$ 0.6 Gyr. Allowing for the 
spread in the Li abundances of the clusters' members, and the 
intrinsically poor ability of Li abundances to discriminate age, this 
result does not actually constrain the age of $\kappa^1$~Cet, but 
suggests that it is very young and not abnormal in its Li depletion 
history.

\subsection{Age}

Rotation periods for \kap have been estimated from spot modulations. The 
determinations range between 8.9 and 9.4 days 
\citep{gudeletal1997,baliunasetal1995,messinaguinan2003,rucinskietal2004}. 
The variation is probably caused by the interplay of differential 
rotation and spots arising at different stellar latitudes of a star 
rotating with $P=8.77$ days in the equator \citep{walkeretal2007}. In 
any case, the rotation period is within the typical range for solar-type 
stars in the Hyades cluster \citep{radicketal1995}, although close to 
the high end \citep{gudeletal1997}. The X-ray luminosity \citep[$\log 
L_{\rm X}=28.8$ in cgs;][]{gudeletal1997} is also comparable to that of 
the Hyades solar analogs \citep{barrado1998}. The same compatibility 
with Hyades members is also encountered in the case of chromospheric 
H$\alpha$ emission and in the abundance of Li in the atmosphere. All 
these indicators seem to suggest an age in the range 0.6--0.8 Gyr. In 
contrast, some age determinations based on chromospheric emission 
indices (like the \ion{Ca}{2} H\&K index $\log R'_{\rm HK}$) seem to 
suggest a younger age around $\sim$ 0.4 Gyr 
\citep{lachaumeetal1999,mamajekhillen2008}.

All the age determination methods employed above have significant 
uncertainties associated. For example, the rotation period vs. age 
relationship may carry a significant cosmic dispersion since not all 
stars have the same initial conditions and the same spin-down 
properties. The Li abundance determination has been observed to have 
nearly one order of magnitude dispersion even within coeval clusters. 
And, finally, chronology with activity indicators may suffer also from 
cosmic dispersion, metallicity effects, etc. In summary, attributing a 
single age value for \kap is difficult and uncertain, but, putting all 
indicators together, we conclude that its age is likely to be in the 
range 0.4--0.8~Gyr (i.e., $0.6 \pm 0.2$~Gyr), which is the value we 
adopt here.

\section{Evolutionary state}

\begin{figure*}
\epsscale{1.0}
\plottwo{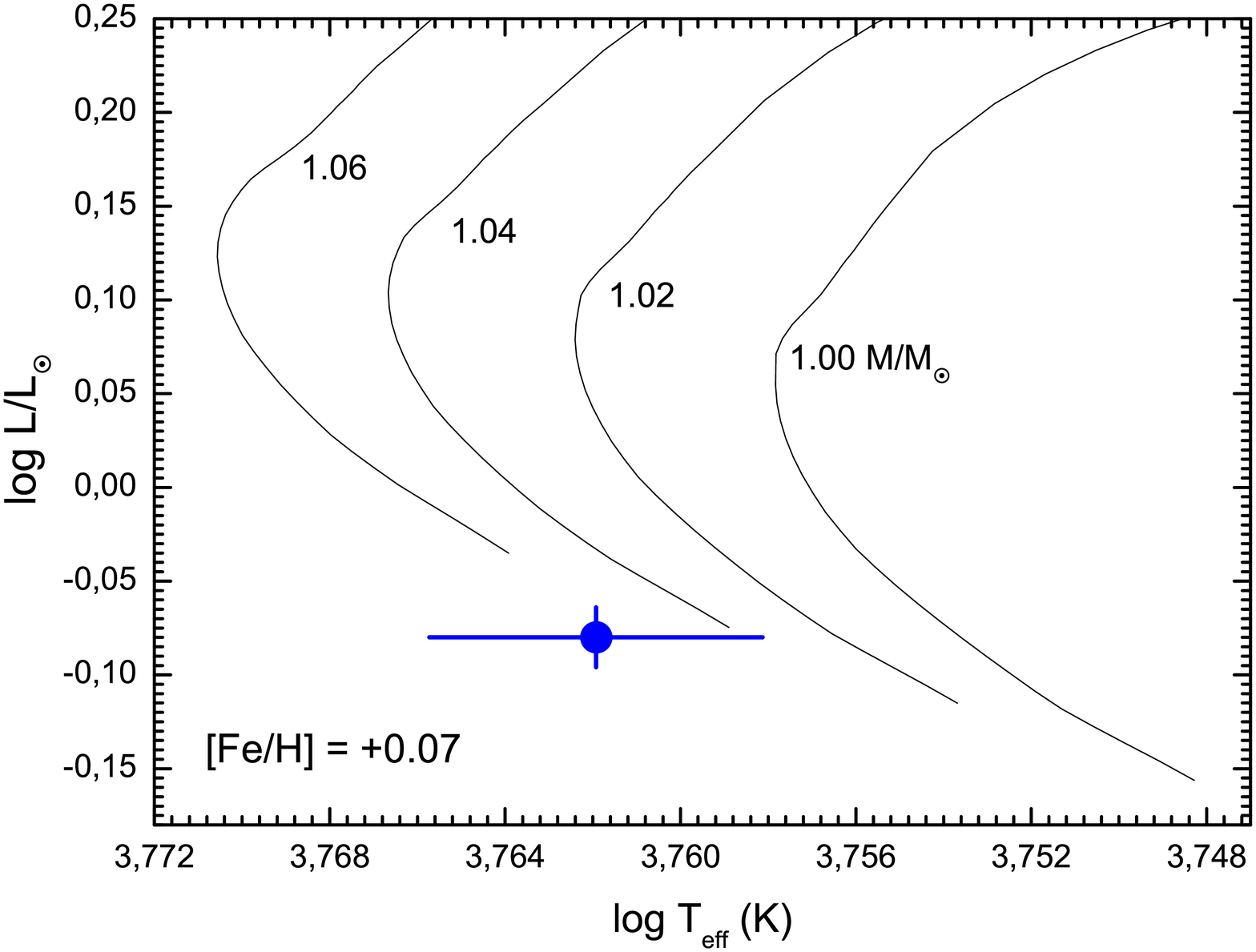}{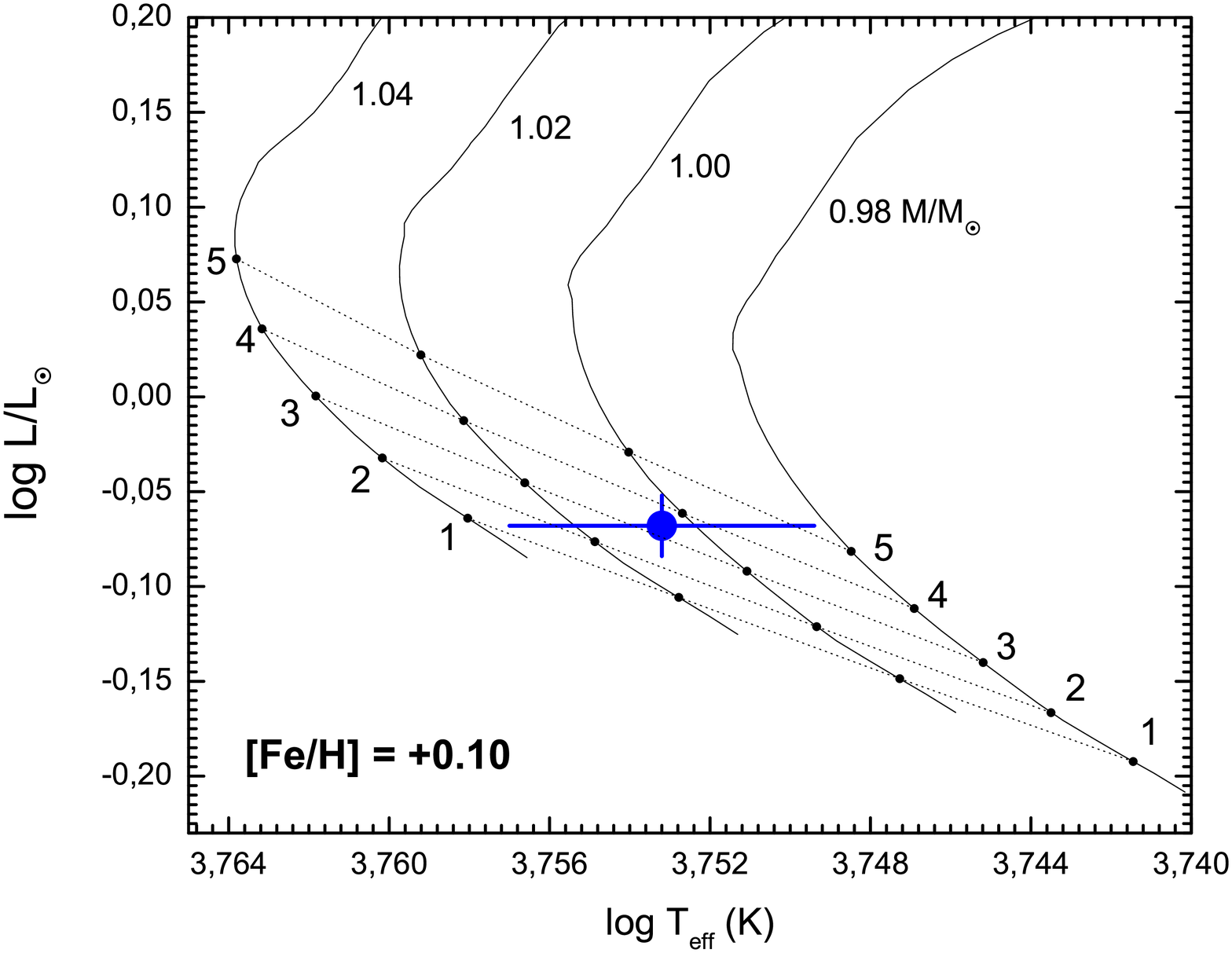}
\caption{{\em Left:} The state of evolution of \kap in a suite of models 
with $[Fe/H]=+0.07$, the spectroscopic metallicity. The star is plotted 
with the spectroscopic \Teff = 5780~K. Evolutionary tracks are labeled in 
solar masses. Note that the diagram does not correspond to solar 
metallicity. {\em Right:} Same as Fig~\ref{HR-feh-spec} for 
$[Fe/H]=+0.10$. Dots along the tracks are labeled by ages in Gyr. Loci of 
same age between the 0.98 and 1.04~M$_{\odot}$ tracks are linked by thin 
lines. \kap is plotted with the average \Teff from photometry and 
H$\alpha$ profile fitting, \Teff = 5665~K.} \label{HR-feh-spec}
\end{figure*}

Given the level of magnetic activity of $\kappa^1$~Cet, it is expected 
that the star still remains very close to, but not exactly on, its 
zero-age main sequence (ZAMS) position. We plotted \kap in two 
theoretical HR diagrams of the Yonsei-Yale 
\citep{kimetal2002,yietal2003} suite of models (Y$^2$). The first 
corresponds to the purely spectroscopic solution of the atmospheric 
parameters (Fig.~\ref{HR-feh-spec} top), \Teff = 5780~K and $[Fe/H] = 
+0.07$, and the second to the photometric/H$\alpha$ determination 
(Fig.~\ref{HR-feh-spec} bottom), \Teff = 5665~K, and the Fe abundance 
from \ion{Fe}{2} lines, $[Fe/H] = +0.10$. The absolute magnitude from 
the Hipparcos parallax ($109.8 \pm 0.78$~mas), coupled to the bolometric 
correction of \citet{flower1996}, results in $\log L/{\rm L}_\odot = 
-0.080 \pm 0.016$, for \Teff = 5780~K, and $\log L/{\rm L}_\odot = 
-0.070 \pm 0.016$, for \Teff = 5665~K (both in a scale in which ${M_{\rm 
bol}}_\odot = 4.75$ and ${BC_{\rm bol}}_\odot = -0.07$). For the \Teff, 
an uncertainty of 50 K was adopted. The diagrams were slightly displaced 
in \Teff and $\log L/{\rm L}_\odot$ (by +0.00186 and +0.0106 in log, 
respectively) so that a solar metallicity and solar mass track reaches 
the position of the Sun exactly at \Teff = 5780~K for 4.53~Gyr 
\citep{guentherdemarque1997}. This is acceptable for our current 
differential study.

To study the evolutionary state of \kap in a more thorough manner we 
performed a series of Monte Carlo simulations taking into account both 
the ``hot'' and ``cool'' scenarios, with their corresponding 
luminosities and metallicities. For this purpose we have generated 
random values ($N=100000$) with a Gaussian distribution for each of the 
input parameters, considering the errors as the standard deviations of 
the distributions. For each combination of the values generated ($T_{\rm 
eff}$, $\log (L/{\rm L}_{\odot})$ and $[Fe/H]$) we carried out an 
interpolation in the synthetic Y$^2$ tracks to obtain the value of the 
mass and the age. Those parameter combinations that yielded unphysical 
situations (i.e., below the ZAMS) were discarded. In the case of the 
``cool'' scenario, 85603 parameter realizations could be used while only 
19110 led to physical solutions in the case of the ``hot'' scenario.

\begin{figure*}
\epsscale{1.0}
\plotone{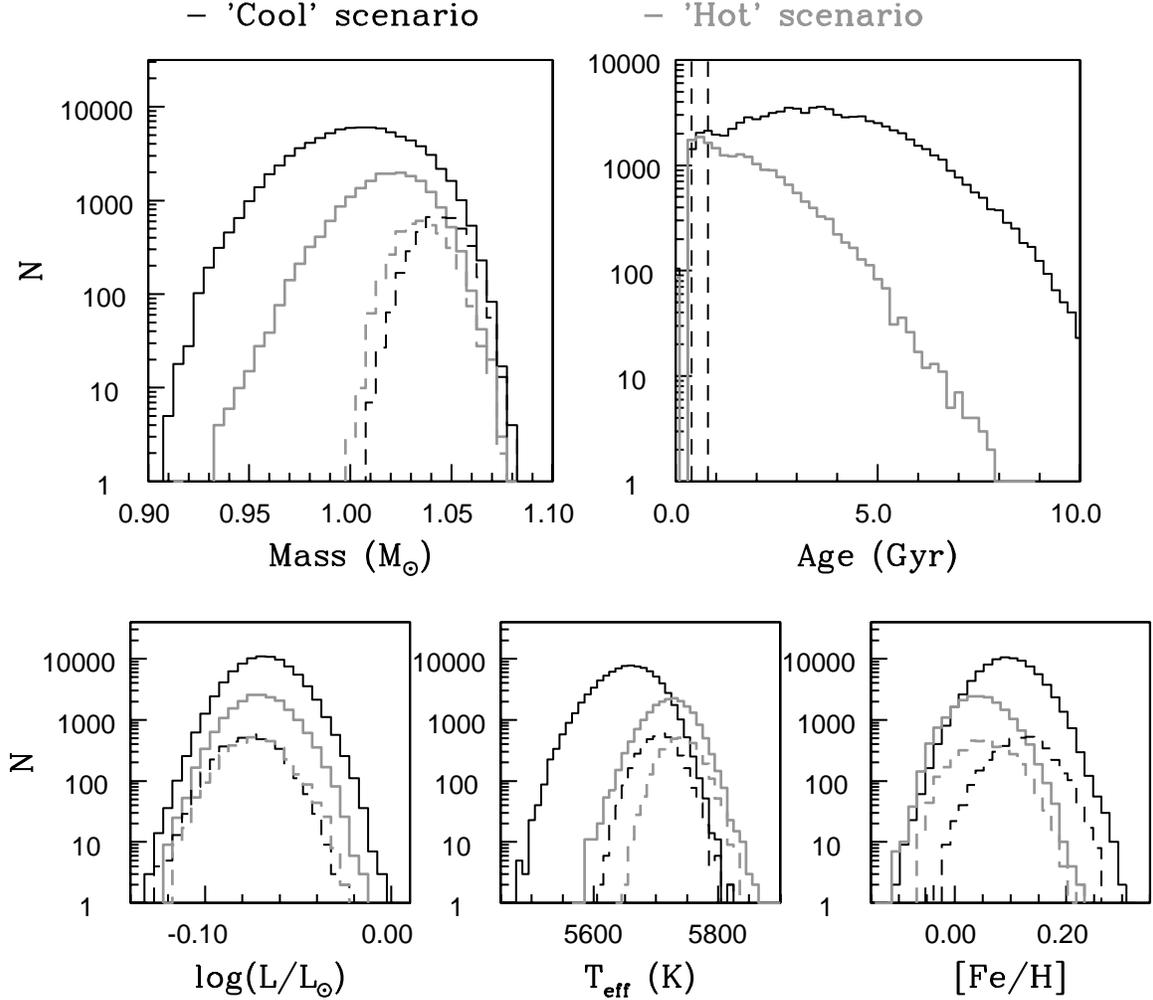}
\caption{Distribution of results of interpolation in the Y$^2$ stellar 
models from $100000$ MonteCarlo realizations using the parameters of the 
``cool'' scenario (i.e., \Teff = 5665 $\pm$ 30~K, $\log L/{\rm L}_\odot 
= -0.070 \pm 0.016$, and $[Fe/H] = +0.10 \pm 0.05$) and the ``hot'' 
scenario (i.e., \Teff= 5780 $\pm$ 30~K, $\log L/{\rm L}_\odot = -0.080 
\pm 0.016$, and $[Fe/H] = +0.07 \pm 0.04$). The top and bottom panels 
correspond to the output (2) and input (3) parameters, respectively. The 
dashed lines represent the distribution of parameters of those solutions 
yielding an age in the interval 0.4--0.8 Gyr.}
\label{Montefig}
\end{figure*}

For our statistical study we only took into account those parameter 
combinations that resulted in an interpolated age in the range 0.4--0.8 
Gyr, which were 4047 and 3715 for the ``cool'' and ``hot'' scenarios, 
respectively. According to this result there is no significantly better 
scenario. In both, we calculated the averages and standard deviations of 
the mass (the average of the age was obviously close to 0.6 Gyr), which 
were found to be $1.045 \pm 0.011$~M$_{\odot}$ and $1.036 \pm 
0.012$~M$_{\odot}$ for the ``cool'' and ``hot'' scenarios, respectively. 
In the case of the input parameters, the average of the solutions 
yielding the correct age are \Teff = 5705 $\pm$ 30~K, $\log (L/{\rm 
L}_{\odot}) = -0.076 \pm 0.014$ and $[Fe/H] = +0.12 \pm 0.05$ for the 
``cool'' scenario, and \Teff = 5740 $\pm$ 30~K, $\log (L/{\rm 
L}_{\odot}) = -0.073 \pm 0.015$ and $[Fe/H] = +0.06 \pm 0.05$, for the 
``hot'' scenario. The luminosities are essentially identical, while the 
temperatures are both pushed into mutual agreement, suggesting that a 
value close to the average is favored by the theoretical models. 
Regarding the metallicity, the behavior is clearly different and the 
model calculations for the ``cool'' and ``hot'' scenarios tend to prefer 
slightly divergent values.  This is expected since the metallicity has a 
strong effect in shifting the theoretical tracks, which, taking into 
account that the \Teff and $[Fe/H]$ uncertainties dominate the error 
budget, naturally constrains the physically acceptable solutions for 
each scenario within a narrow range of metallicity. Fig. \ref{Montefig} 
shows the histograms of the input and output parameters for the ``cool'' 
and ``hot'' scenarios. The solid lines depict the distribution from the 
entire simulation while dashed lines show the distribution of those 
solutions that yield an age in the interval 0.4--0.8 Gyr.

The simulations favor slightly the ``cool'' spectroscopic solution and, 
although the statistical significance of this result is low, when 
considered with the likely presence of non-local radiative fields 
distorting the \ion{Fe}{1} and \ion{Fe}{2} populations, we find the 
``cool'' spectroscopic solution more consistent and adopt it in the 
subsequent discussion.

The position of \kap in the theoretical HR diagram, along with its 
magnetic activity context, is thus compatible with a slightly metal-rich 
star, slightly more massive than the Sun, and $\sim$ 0.4--0.8 Gyr old. 
These \Teff and $\log L$/L$_\odot$ values, along with $M = 1.04$~M$_\odot$, 
when put into the well known equation:
\[ \log\Bigg( \frac{g}{g_{\odot}}\Bigg) = \log\Bigg( \frac{M}{M_{\odot}}\Bigg)
+ 4 \log\Bigg( \frac{T_{\mbox{eff}}}{T_{\mbox{eff}\odot}}\Bigg) -
\log\Bigg( \frac{L}{L_{\odot}}\Bigg)\] 
lead to $\log g = 4.49$, in excellent agreement with the spectroscopic 
solution, and confirming that \kap is very close to the ZAMS. This surface 
gravity estimate is very insensitive to uncertainties in the inferred 
mass.

Considering the Galactic orbit of \kap in the context of nearby 
solar-type stars as analyzed by \citet{portodemelloetal2006}, \kap is 
seen to have one of the lowest orbital eccentricities in the local 
population, and a mean galactocentric radius nearly identical to that of 
the Sun, and therefore a very similar Galactic orbit.

\section{UV Irradiance}

Because of its apparent brightness, \kap has been subject to intensive 
scrutiny with a variety of X-ray and UV telescopes \citep[Paper 
I]{gudeletal1997,telleschietal2005} that have revealed fluxes 
significantly higher than those of the current Sun. Paper I carried out 
an analysis of high-energy data with a wide wavelength coverage and 
found X-ray and EUV fluxes for \kap some 20 and 10 times stronger, 
respectively, than today's Sun. In the FUV and UV, beyond the H Lyman 
$\alpha$ line (121.5 nm) only fluxes for some strong features have been 
presented, mostly from IUE spectra \citep[Paper I]{A97}, but not the 
needed overall irradiance (strong lines + pseudo-continuum).

We have compiled flux data of \kap in the UV and FUV. Spectra covering 
from 93 to 118~nm are available from FUSE and were discussed by 
\citet{guinanetal2003} and in Paper I. Here we have used only night-time 
data since the relevant H Lyman lines are prone to contamination from 
geocoronal emission. The interstellar medium-corrected profile of the 
strong H Ly $\alpha$ line was taken from \citet{woodetal2005}, which 
agrees well with the values provided in Paper I. The rest of the 
spectrum (i.e., from H Ly$\alpha$ up to nearly 300 nm) was taken from 
the CoolCAT 
catalog\footnote{http://casa.colorado.edu/$\sim$ayres/CoolCAT/}, which 
compiles echelle spectroscopy obtained with the HST STIS instrument. The 
combined spectrum, covering from 93 to 299~nm (with a few small gaps), 
was smoothed using a convolution with a gaussian of 0.3 nm FWHM and 
resampled at steps of 0.01~nm. The random uncertainties after such 
convolution are negligible above 200 nm (well below 1\%) and then slowly 
increase towards shorter wavelengths to reach about 10\% at 140 nm and 
20\% at 93 nm. At long wavelengths the dominant error will be that of 
the standard flux calibration but, according to STIS specifications and 
the information in CoolCAT, the systematic uncertainty should not exceed 
5\%.

\begin{figure*}
\epsscale{1.0}
\plottwo{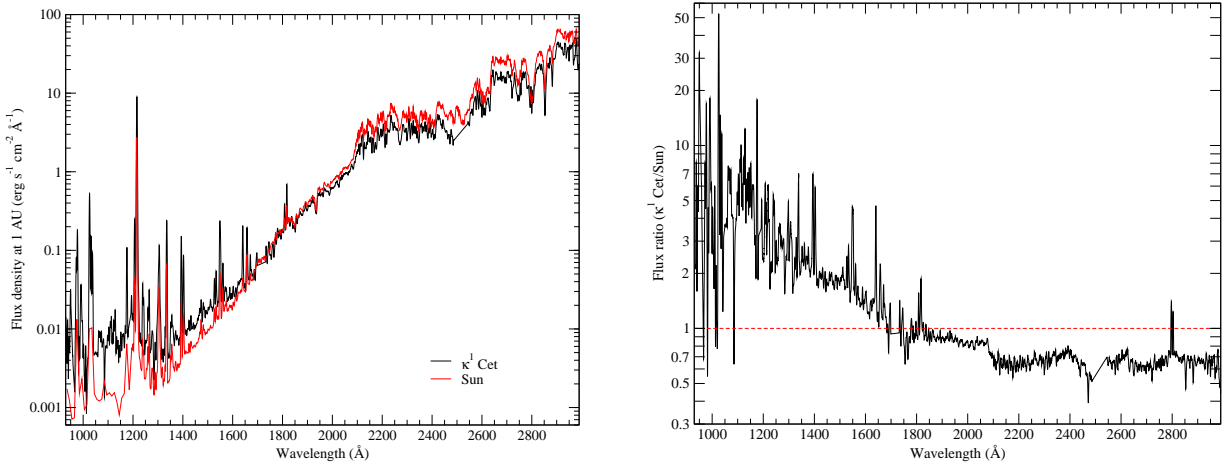}{ribas_f07b.eps}
\caption{{\em Left:} Comparison of the observed UV spectra of \kap
and the current Sun. {\em Right:} Ratio of the observed UV spectra
of \kap and the current Sun.} \label{fig:comp}
\end{figure*}

The \kap spectrum was compared with a spectrum of the Sun obtained from 
\citet{thuillieretal2004}, corresponding to medium solar activity. Fig. 
\ref{fig:comp} illustrates such comparison. The left panel shows the 
normalized flux density at a distance of 1 AU from both \kap and the 
current Sun, while the right panel depicts the normalized flux ratio. 
The results clearly show that \kap is about 35\% fainter than today's 
Sun for wavelengths above 210 nm, a range completely dominated by 
thermal radiation (except for the chromospheric emission of the 
\ion{Mg}{2} h\&k lines). Below 210 nm the difference in the ratio 
decreases to about 17\% at 200 nm, 10\% at 190 nm and the fluxes equal 
on average around 170 nm. Some chromospheric lines start to appear below 
185 nm and those are much stronger in the case of $\kappa^1$~Cet. 
Shortward of 170 nm the relative flux of \kap increases up to about a 
factor of 4 at 120 nm and then even higher below. At those wavelengths 
most of the flux is dominated by chromospheric lines that are 
significantly stronger for $\kappa^1$~Cet.

The results in Fig. \ref{fig:comp} clearly illustrate how the Sun, 
because of its higher temperature, possesses stronger photospheric UV 
radiation, yet, when the chromosphere emission is considered, \kap 
emissions are stronger because of its higher level of magnetic activity. 
The transition between photosphere and chromosphere in terms of emission 
is usually set at about 170 nm. However, note how in the interval 
between 170 nm and 210 nm, \kap is relatively brighter than expected. 
The flux ratio difference with the Sun is about 10--15\% while about 
35\% is found at longer wavelengths.

\citet{cnossenetal2007} presented estimates of the UV flux for 
$\kappa^1$~Cet. The authors did not use measured UV data but employed 
astrophysical plasma models to generate synthetic flux distributions 
from an emission measure distribution (EMD). Such EMD was obtained from 
the analysis of X-ray and EUV data, typically arising from coronal 
emission. The calculated UV fluxes, originating from much cooler plasma 
in the chromosphere, should thus be a crude approximation of reality. 
Comparison of the fluxes obtained by \citet{cnossenetal2007} is not 
possible beyond a general apparent agreement, but real data, also with 
much higher resolution, ought to be preferred.

\section{A Young Sun}

\begin{figure}
\epsscale{1.0}
\plotone{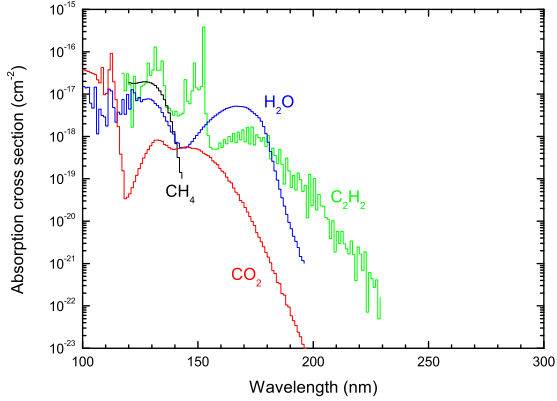}
\caption{Photoabsorption cross sections of some molecules suspected to
have been present in early Earth's atmosphere.} \label{cross-sections}
\end{figure}

Evidence presented here for the young solar analog \kap suggests that 
the Sun was significantly more active in its past. This early activity 
for Sun-like stars is known in the X/EUV range and its evolution with 
the age of the star was studied in Paper I. Interestingly, \kap also 
exhibits enhanced fluxes in the $\sim$ 100--200 nm range. These 
emissions may have had an impact on the early evolution of Earth's 
atmosphere and may have played a role in the origin and development of 
life on Earth: it has been proposed, for instance, that UV radiation may 
help the synthesis of complex ribonucleotides in plausible early Earth 
conditions \citep{powneretal2009}. In this wavelength range, the solar 
emission drives the photochemistry and thus the molecular composition of 
planetary atmospheres. This can be gauged in Fig. \ref{cross-sections}, 
where we show UV photoabsorption cross sections of relevant molecules.

\begin{figure}
\epsscale{1.0}
\plotone{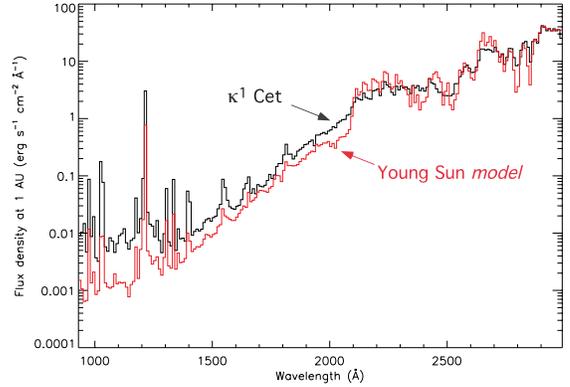}
\caption{The observed $\kappa^1$ Cet spectrum vs. a simple model of 
the young Sun (see text).} \label{best-guessed}
\end{figure}

To illustrate how the spectral irradiance in the 100--300~nm domain can 
influence the photochemistry, we have computed the photolysis rates in 
an early Earth atmosphere subject to two different irradiance spectra. 
The first spectrum is that of $\kappa^1$~Cet. The second spectrum 
(``model'') is an educated guess of the UV spectral irradiance of the 
Sun at 0.6~Gyr based on our knowledge of stellar evolution and spectral 
synthesis, prior to the latests results of the Sun in Time program in 
Paper I, and on the basic assumption that the solar activity has not 
changed during the last 3.9~Gyr.  This model of the young Sun spectrum 
is obtained by adding a synthetic photospheric irradiance spectrum and 
an activity component. The synthetic spectrum was computed by F. 
Castelli with the ATLAS9 model for the characteristics of the Sun at 0.6 
Gyr ($L=0.73$~L$_\odot$, \Teff = 5680~K), as derived from the Y$^2$ 
stellar evolution models. The short wavelength emission associated with 
activity is assumed to be that of today's Sun. Such activity component 
can be calculated as the difference between the observed spectrum 
\citet{thuillieretal2004} and the photospheric synthetic 
spectrum\footnote{ATLAS9 spectra of the present Sun are available at 
http://wwwuser.oat.ts.astro.it/castelli/sun.html}. The difference, 
namely the activity component, becomes significant below about 200~nm 
and dominates below 175~nm. Both spectral irradiances are scaled to 
1~AU. The spectra of this simple young Sun model and \kap are compared 
in Fig.~\ref{best-guessed}.

\begin{figure*}
\epsscale{1.0}
\plottwo{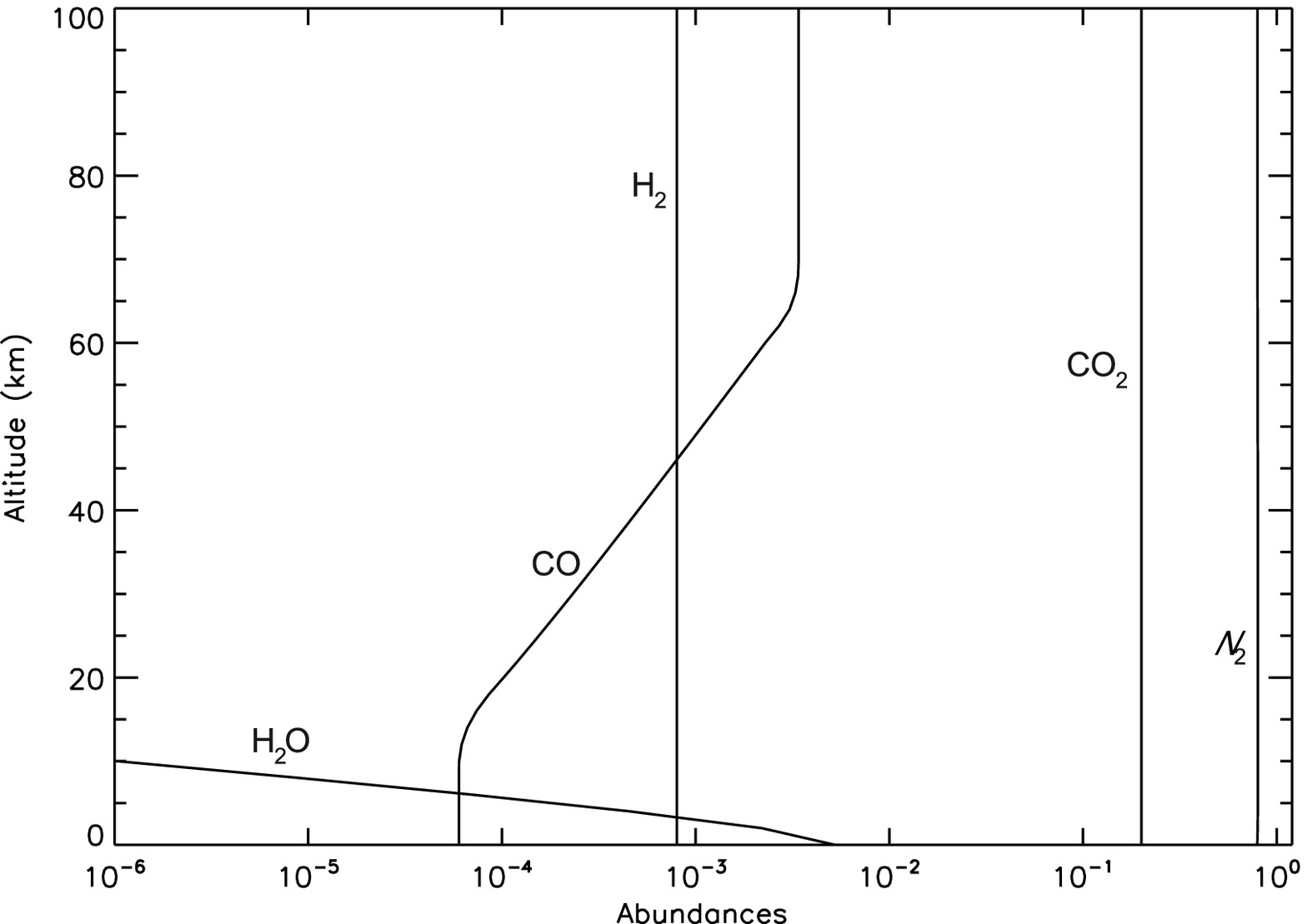}{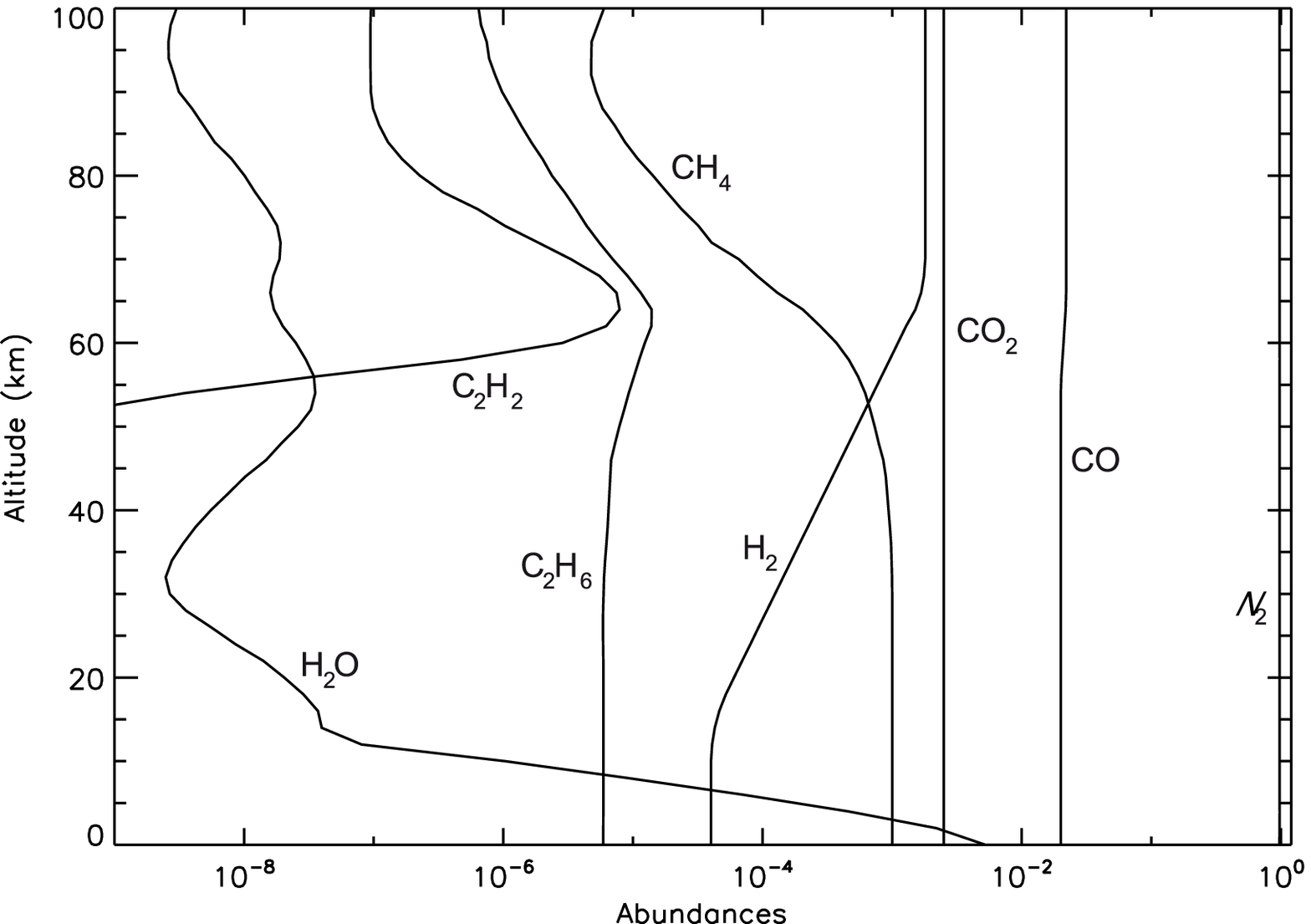} 

\plottwo{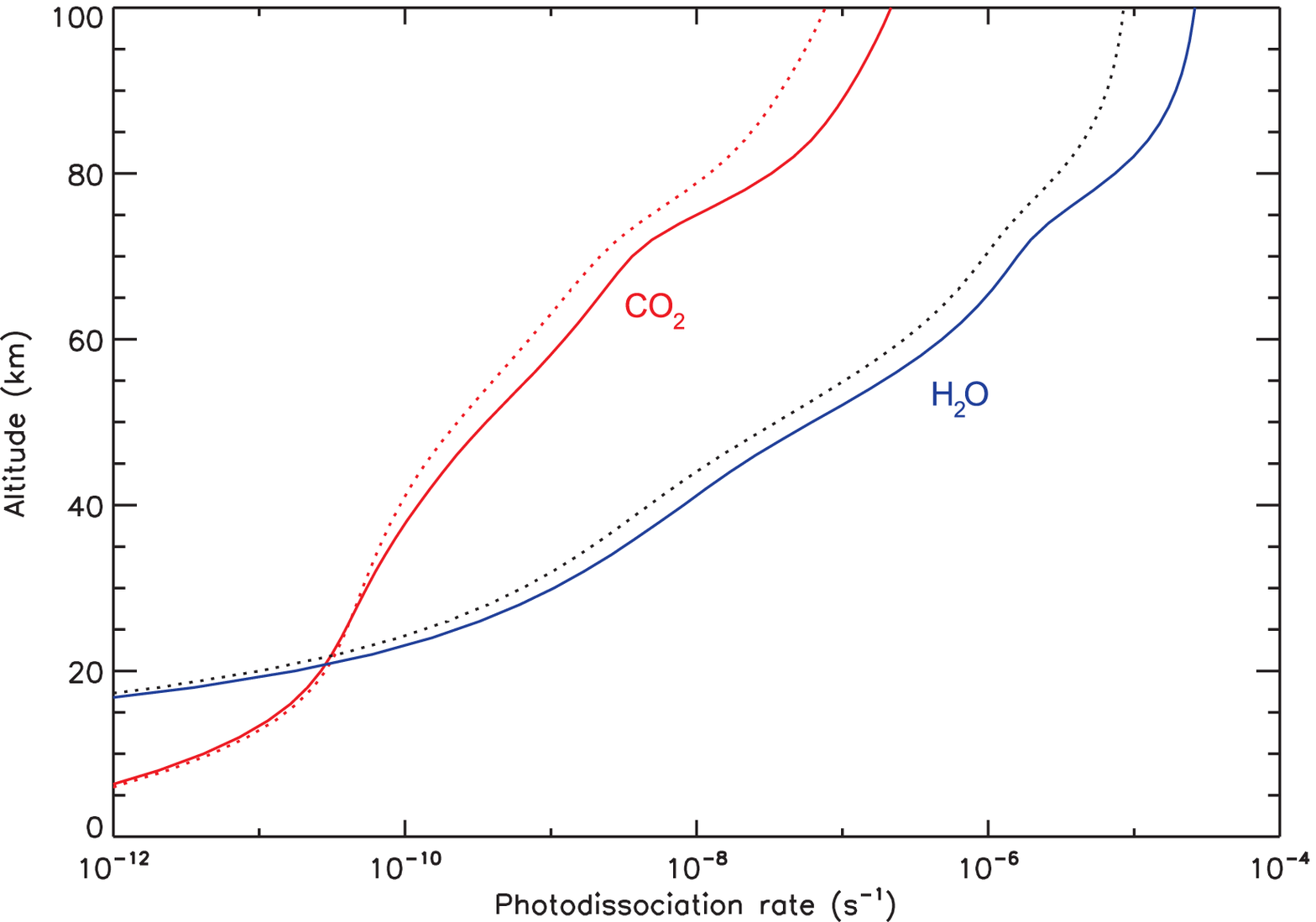}{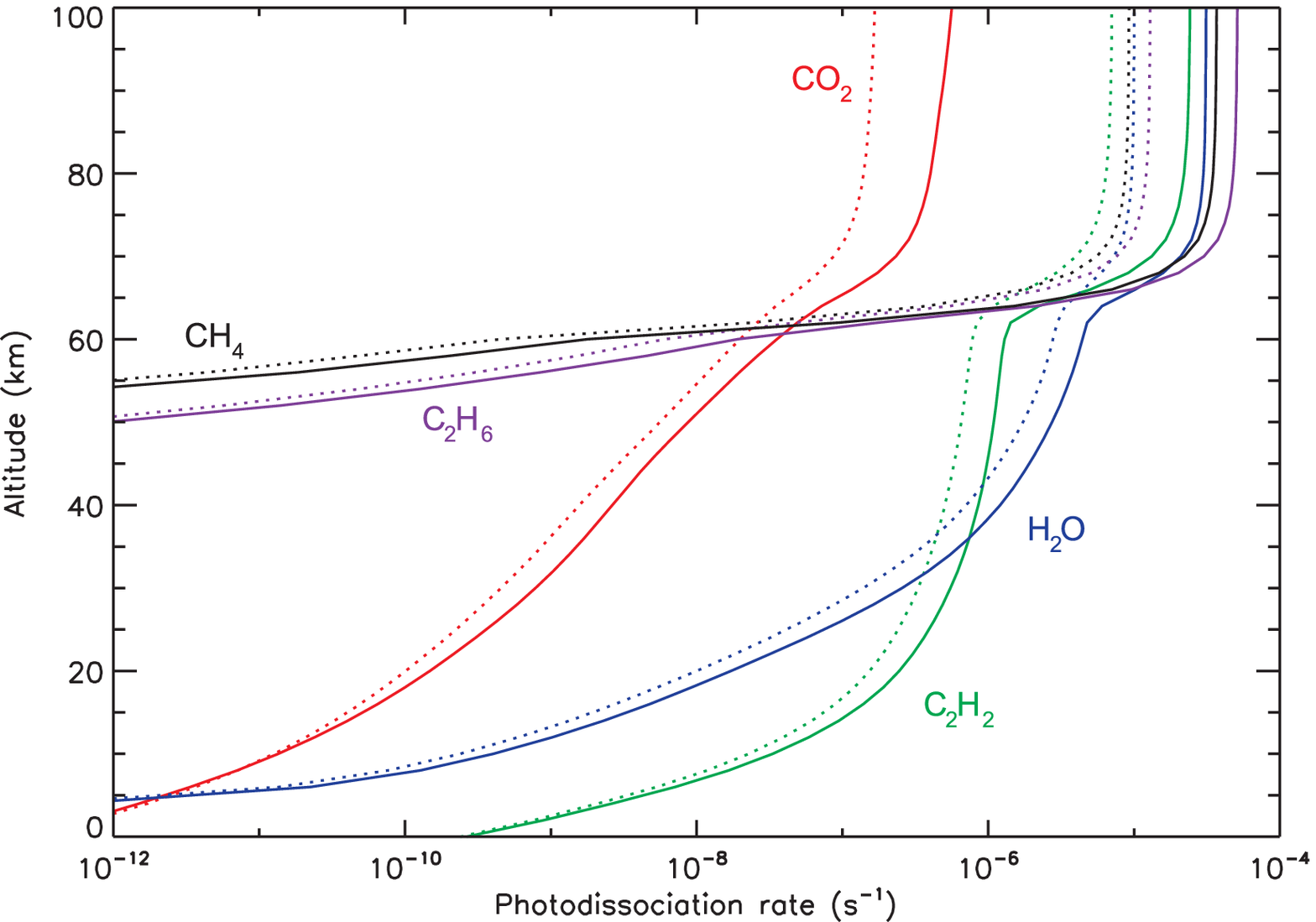} 

\plottwo{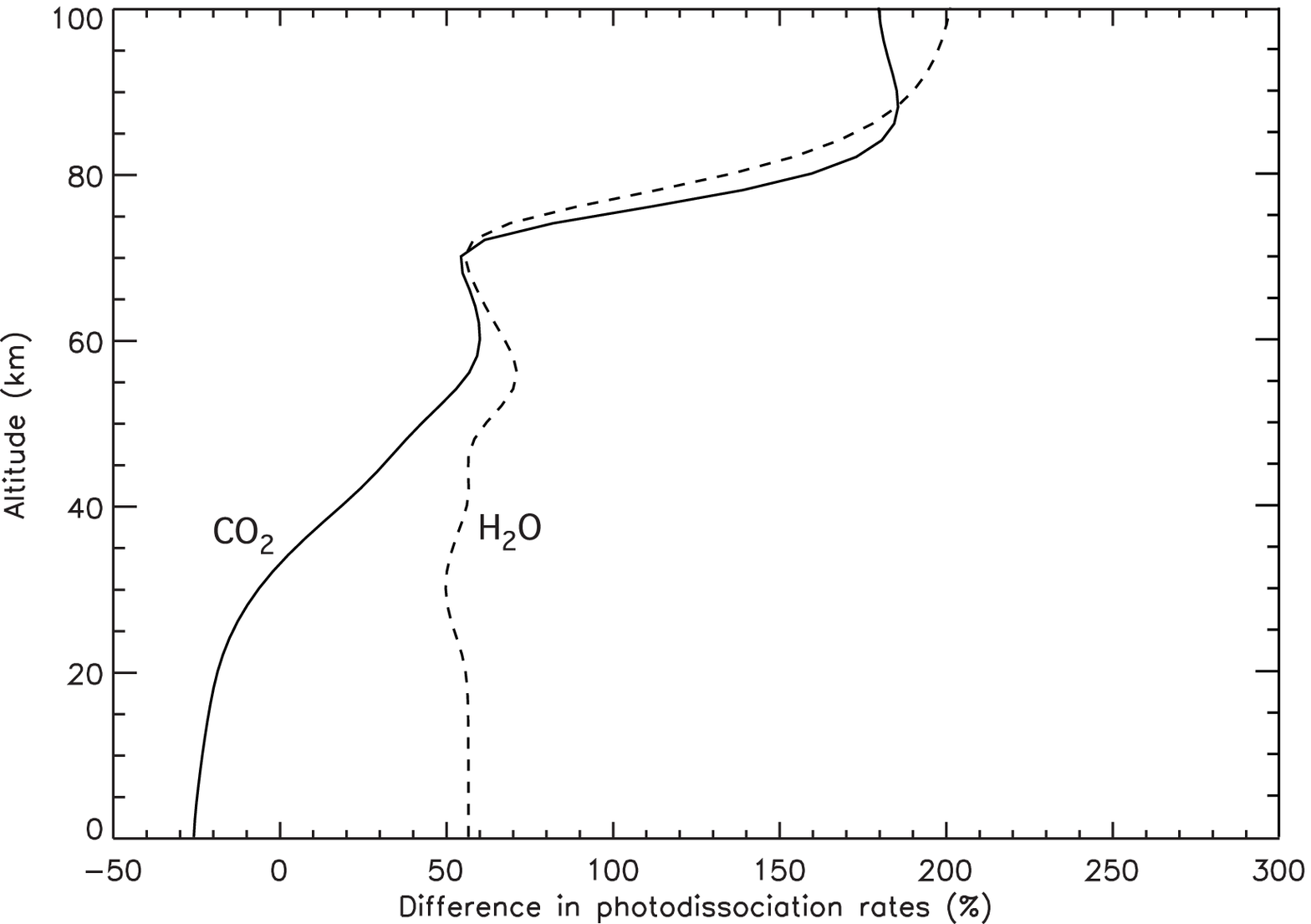}{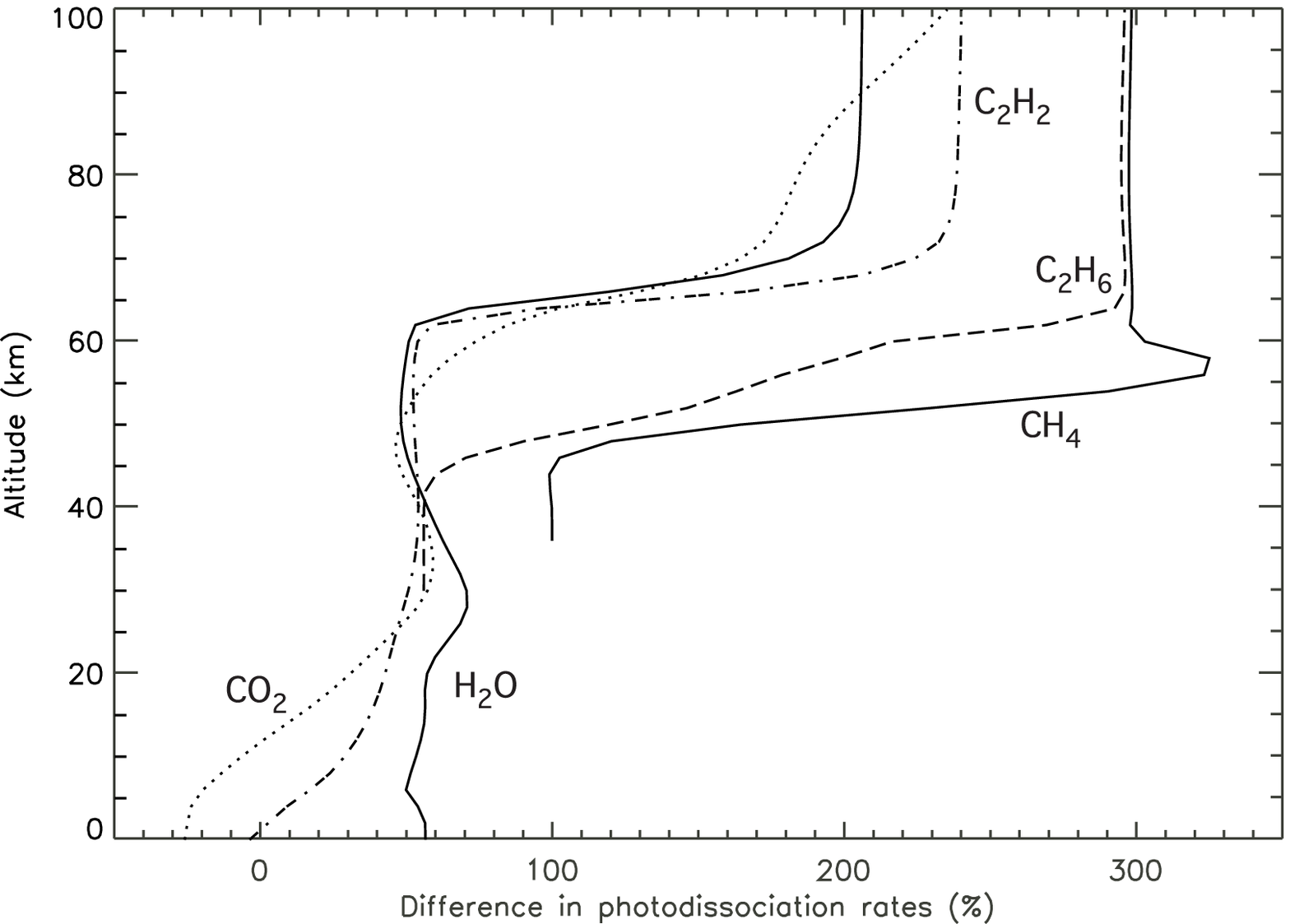} 
\caption{{\bf Top:} The two early Earth atmospheres considered. {\em 
Left:} The ``standard'' atmosphere from \citet{kasting1993}. {\rm Right:} A 
more reducing composition, from \citet{pavlovetal2001}. {\bf Middle:} The 
corresponding photodissociation rates of relevant molecular species 
computed for two different input solar spectra: \kap (solid 
lines) and a theoretical young Sun (dashed lines). {\bf Bottom:} The 
percentage of enhancement in the photodissociation rates when adopting the 
spectral irradiance of $\kappa^1$~Cet.} \label{photochem}
\end{figure*}

We have considered two different atmospheric compositions (see 
Fig.~\ref{photochem}, top):
\begin{itemize}
\item A CO$_2$-rich atmosphere described in \citet{kasting1993}. In this 
model the enhanced level of CO$_2$ provides enough greenhouse warming to 
compensate for the faint luminosity of the Sun 3.9~Gyr ago. The abundance 
of H$_2$ is obtained for the current volcanic emission and assuming a 
diffusion-limited rate for the escape of hydrogen to space. This 
atmosphere contains no organic species.
\item A more reduced atmosphere excerpted from \citet{pavlovetal2001}, that 
differs from the previous one by a significant level of CH$_4$ and some 
photochemically produced hydrocarbons (in particular C$_2$H$_2$ and 
C$_2$H$_6$). In \citet{pavlovetal2001}, the source of CH$_4$ is assumed to 
be biogenic methanogenesis, but abiotic sources associated with 
hydrothermal activity are also plausible 
\citep{sleepetal2004,albaredeetal2009}.
\end{itemize}

Both compositions include 1~bar of N$_2$ as the main constituent. For 
both cases, our atmospheric thermal profile has been deduced from a 
present day terrestrial profile, assuming a constant stratospheric 
temperature equal to 200~K between 14 and 60~km. This is of course not 
consistent with the detailed compositions but has little impact on the 
photolysis rates. Also, the detailed composition from 
\citet{kasting1993} and \citet{pavlovetal2001} were computed with a 
photochemical model that assumed different UV fluxes. The purpose here 
is not to run a consistent photochemical model, which will be the done 
in future studies, but to illustrate how the photodissociation rates for 
these atmospheric compositions are sensitive to an enhanced UV 
irradiance.

Photodissociation rates $J_i(z)$ (s$^{-1}$) at the altitude $z$ of the 
different absorbing species $i$ included in the model were computed in the 
range of wavelength $[\lambda_1,\lambda_2]$ as
\begin{equation}
  J_i(z)=\sum_j\left(\int_{\lambda_1}^{\lambda_2}\!\!\!q_{i,j}(\lambda)\sigma_{i}(\lambda)F(\lambda,z)d\lambda\right)
\end{equation}
which requires beforehand the implementation of their absorption 
cross-sections $\sigma_{i}(\lambda)$, of their different photodissociation 
pathways $j$ each characterized by a quantum yield $q_{i,j}(\lambda)$ and 
of the incident stellar UV flux at every level in the atmosphere 
$F(\lambda,z)$ as well. This incident stellar flux $F(\lambda, z)$ was 
calculated as a function of the diurnally averaged unattenuated stellar 
flux at the top of the atmosphere $F_\infty(\lambda)$ by considering only 
the integrated molecular absorption and Rayleigh scattering within 
molecular nitrogen N$_2$.
\begin{equation}
F(\lambda,z) = F(\lambda,\infty) e^{-\tau_{\rm abs+diff}(\lambda,z)}
\end{equation}

The photolysis rates computed for both atmospheric compositions and both 
UV spectral irradiances are shown in Fig.~\ref{photochem}. These rates 
are found to be enhanced by a factor of 2--3 in the stratosphere and 
mesosphere for H$_2$O and CO$_2$ and by a factor of 4 for organic 
compounds. These photodissociations produce the radicals that activate 
the photochemical evolution, which is controlled by several hundreds of 
individual reactions, coupled with the vertical mixing and the 
condensation of some species.  The resulting set of equations behaves 
strongly non-linearly and can be very sensitive to the photolysis rates, 
which is why the UV spectrum has a critical influence. Therefore, 
considering \kap as a close analog of our Sun in its early times, such 
enhanced photodissociation rates were likely to trigger a peculiar 
atmospheric chemistry that we plan to investigate shortly by using a 
photochemical model of the primitive Earth's atmosphere. The higher 
H$_2$O photolysis, associated with the intense heating of the 
thermosphere by the strong EUV irradiance, would also result in a higher 
escape rate of hydrogen to space, particularly important for the early 
evolution of Mars and Venus. The effect of the relatively strong UV 
fluxes is thus still significant for a 0.6 Gyr old Sun-like star, and is 
likely to play an even more important role for younger stars, and for 
the atmospheric processes that occurred during the earliest stages of 
our planet.

At wavelengths above 200~nm, the photospheric emission dominates and the 
spectrum can be computed using a stellar atmosphere model. However, the 
inaccuracy of synthetic spectra can have a significant effect on the 
photodissociation rates. This inaccuracy has various origins as, for 
instance, the opacities, the elemental abundances, the limb 
darkening/brightening. \citet{edvardsson2008} compared different models 
with observed stellar spectra, above 300~nm, and reported the occurrence 
of numerous patterns with widths of 0.5-1.5~nm, below 450~nm, with 
systematic uncertainties of about 10\%. We noted even higher differences 
below 300~nm when comparing synthetic spectra obtained for the same 
stellar parameters but with different models. We can see in 
Fig.~\ref{best-guessed} that above 200~nm the synthetic model used for 
the young Sun fluctuates around that of $\kappa^1$~Cet. This difference 
can be attributed to both the inaccuracy of the model and to the fact 
that \kap is a close but not exact replica of the Sun, having a slightly 
different mass and metallicity.

Between 150 and 200 nm, the photon flux increases by more than two 
orders of magnitude. Which means that a 10\% error on a 1~nm bin at 
200~nm represent 10 times more photons than all the flux in a 1~nm bin 
at 150~nm. We can thus wonder about the influence of this error on the 
photodissociation rates. In the atmosphere models we considered, most of 
the species are photodissociated at wavelengths below 200~nm and most 
absorption cross-sections decrease by orders of magnitude between 150 
and 200~nm. As a consequence, variations of the flux above 200~nm have a 
limited impact on the dissociation rates. For some species that absorb 
at $\lambda > 200$~nm (like CO$_2$ or hydrocarbons), photolysis rates 
can be slightly higher in the lowest atmosphere when computed with the 
young Sun model. But with the atmospheric compositions we considered, we 
checked that this effect remains negligible by comparing the rates from 
Fig.~\ref{photochem} with rates computed when both spectra are set to 
the same values at wavelengths above 200~nm. We should stress however 
that other atmospheric compositions, involving species that have strong 
absorption cross-sections at $\lambda > 200$~nm (like for instance 
SO$_2$ or O$_3$), would be much more sensitive to this part of the 
spectrum. This demonstrates that using both the observed spectrum of a 
young Sun proxy and a theoretical model may be necessary for a full 
description of planetary atmospheres.

\section{Summary and conclusions}

In this paper we have carried out an in-depth study of the bright, 
nearby solar analog $\kappa^1$~Cet. Several methods have been used to 
estimate its effective temperature and chemical composition, yielding 
preferred values of \Teff = 5665 $\pm$ 30~K, and $[Fe/H] = +0.10 \pm 
0.05$. The systematic offset between \Teff values obtained from 
photometry/line profiles and the excitation/ionization \ion{Fe}{1} and 
\ion{Fe}{2} equilibria is evidence of non-LTE effects, probably related 
to UV overionization due to the strong magnetic activity of this star. 
Adopting as the best atmospheric parameters the photometry and H$\alpha$ 
profile \Teff and the \ion{Fe}{2} metallicity, we have been able to set 
constraints to the stellar age, which should be between 0.4 and 0.8~Gyr. 
All the information gathered indicates that \kap is a star with nearly 1 
solar mass in a relatively unevolved evolutionary stage. As such, it is 
an excellent match to the Sun as it was some 3.7--4.1 Gyr ago.

The radiation from the young Sun must have played an essential role in 
shaping the atmospheres of the Solar System planets. In particular, the 
UV flux is responsible for the photochemical processes in the 
atmosphere. We have been able to compile data, taken both with FUSE and 
HST, covering the entire UV and it shows that $\kappa^1$~Cet's flux is 
some 35\% lower than the current Sun's between 210 and 300 nm, it 
matches the Sun's at 170 nm and increases to at least 2--7 times higher 
than the Sun's between 110 and 140~nm. We have compared these fluxes 
with a ``theoretical'' young Sun estimated by adding the current 
chromospheric flux to a photospheric model with the correct radiative 
properties. We have used a photochemical model to calculate the 
photodissociation rates of the most relevant molecules in the assumed 
composition of early Earth's atmosphere. The results indicate that such 
rates should have been several times higher than those resulting from a 
simplistic ``theoretical'' solar spectrum.

Our calculations demonstrate that self-consistent planetary atmosphere 
calculations must account for the much stronger photodissociating 
radiation of the young Sun. The resulting chemistry could be 
significantly different from that commonly assumed. This is obviously 
very relevant at a significant point in the Solar System evolution, when 
life was gaining a secure foothold on Earth and Mars lost its liquid 
water inventory.

\acknowledgments

G. Thuillier is thanked for providing the high-resolution solar 
spectrum. T. Ayres is thanked for making the HST/STIS spectra available 
via the CoolCAT catalog. The referee is thanked for a number of useful 
comments that have led to an improved paper. IR, SC and AG acknowledge 
support from the Spanish Ministerio de Ciencia e Innovaci\'on via grant 
AYA2006-15623-C02-01. GFPM acknowledges financial support by CNPq grant 
n$^{\circ}$ 476909/2006-6, FAPERJ grant n$^{\circ}$ 
APQ1/26/170.687/2004, and a CAPES post-doctoral fellowship n$^{\circ}$ 
BEX 4261/07-0. LDF thanks CAPES for a MSc scholarship. FS acknowledges 
support from the European Research Council (Starting Grant 209622: 
E$_3$ARTHs). EH acknowledges support by a post-doctoral fellowship 
funded by the Conseil R\'egional d'Aquitaine and by the Fondation Louis 
D. SC is supported by a Marie Curie Intra-European Fellowship within the 
7th European Community Framework Programme. We thank the staffs of 
OPD/LNA and ESO for support in the observing runs performed for this 
project. Use was made of the Simbad database, operated at CDS, 
Strasbourg, France, and of NASA's Astrophysics Data System Bibliographic 
Services.

\end{document}